\pdfoutput=1

\documentclass[11pt]{article}
\usepackage{amssymb}
\usepackage{graphicx}
\parskip \baselineskip

\newtheorem{lm}{Lemma}

\newtheorem{prop}{Proposition}

\newtheorem{corol}{Corollary}

\newtheorem{ex}{Example}

\newtheorem{re}{Remark}

\newcommand{\ol}{\overline}

\date{}

\begin{document}

\title{ALLSAT compressed with wildcards: An invitation for C-programmers}

\author{Marcel Wild}

\maketitle

\begin{quote}
{\bf Abstract} The model set of a general Boolean function in CNF is calculated in a compressed format, using wildcards.  This novel method can be explained in  very visual ways.  Preliminary comparison with existing methods (BDD's and ESOPs) looks promising but our algorithm begs for a C encoding which would render it comparable in more systematic ways.
\end{quote}

\section{Introduction}

By definition for us the {\it ALLSAT} problem is the task to enumerate all models of a Boolean function $\varphi=\varphi(x_1,...,x_t)$. In our article $\varphi$ is given by a CNF $C_1\wedge\ldots C_s$ with clauses $C_i$. The Boolean functions can be of a specific kind (e.g. Horn formulae), or they can be general Boolean functions.  The article in front of you is one in a planned series\footnote{Article [W] contains a tentative account of the
planned topics in the series, and it reviews wildcard-related previous publications of the author. The appeal of the article in your hands is its no-fuzz approach (for Theorems look in [W]) and its strong visual component.}  of articles dedicated to the general theme of 'ALLSAT compressed with wildcards'.

While much research has been devoted to SATISFIABILITY, the ALLSAT problem commanded  less attention. The seemingly first systematic comparison of half a dozen methods is carried out in the article of Toda and Soh [TS]. It contains the following, unsurprising finding. If there are billions of models then the algorithms that put out their models one-by-one, stand no chance against the only competitor offering compression. The latter is a method of Toda (referenced in [TS]) that is based on Binary Decision Diagrams (BDD); see [K] for an introduction to BDD's. Likewise the method propagated in the present article has the potential for compression. Whereas BDDs achieve their compression using the common don't-care symbol $\ast$ (to indicate bits free to be 0 or 1) our method employs three further kinds of wildcards, and is entirely different from BDDs. Referring to these wildcards we call it the $men$-algorithm. In a nutshell, the $men$-algorithm retrieves the model set $Mod(\varphi)$ by {\it imposing} one clause after the other:

(1)\quad $\{0,1\}^t\supseteq Mod(C_1)\supseteq Mod(C_1\wedge C_2)\supseteq \cdots\supseteq Mod(C_1\wedge\ldots\wedge C_n)=Mod(\varphi)$

The Section break up is as follows. In Section 2 we visualize the core maneuver for achieving (1). It will turn out that the intermediate stages of shrinking $\{0,1\}^t$ to $Mod(\varphi)$ do not exactly match the $n+1$ 
idealized stages $Mod(C_1\wedge\ldots\wedge C_k)$ in $(1)$.

 Section 3 starts with a well-known Boolean tautology, which for $k=2$ is
$x_1\vee x_2\leftrightarrow x_1\vee(\ol{x_1}\wedge x_2)$. Generally the $k$ terms to the right of $\leftrightarrow$ are mutually exclusive, i.e. their model sets are {\it disjoint}. The problem of keeping systems $r_i$ of bitstrings  disjoint upon imposing clauses on them, is the core technical difficulty of the present article. It will be handled by wildcards that adapt well to the above tautology. While in Section 3 only positive, or only negative clauses are considered (leading to dual kinds of wildcards), both kinds occur {\it together} in Section 4. This requires a third type of wildcard, which in turn makes the systems $r_i$ more intricate. Fortunately (Section 5) this doesn't get out of hand. Being able to alternately impose positive clauses like
 $x_1\vee x_2$ and  negative clauses like $\ol{x_3}\vee\ol{x_4}\vee\ol{x_5}$, does not enable us to impose the {\it mixed} clause  
$x_1\vee x_2\vee\ol{x_3}\vee\ol{x_4}\vee\ol{x_5}$. But it certainly helps (Section 6).
After the brief technical Section 7, in Section 8 we carry out  the $men$-algorithm on some random moderate-size Boolean functions, and observe that the compression achieved compares favorably to BDD's and ESOP's. We calculate the latter two by using the commands {\tt expr2bdd} of Python and {\tt BooleanConvert} of Mathematica. Of course only systematic\footnote{The $men$-algorithm awaits implementation in either high-end Mathematica-code or in C. As to Mathematica, this remains the only programming language I master. If any reader wants to implement in C the $men$-algorithm, e.g. as a PhD topic, then he/she is welcome to seize  this offer on a silver platter. The benefit (as opposed to pointless coding efforts with Mathematica) is that the $men$-algorithm coded in C or C+ becomes comparable to the methods evaluated in [TS], and possibly others.}
experiments will show the precise benefits and deficiencies of the three methods.

\section{ Visualization of the LIFO-stack and the Core Maneuver}

{\bf 2.1} For the time being it suffices to think of a {\it 012men-row} as a row (=vector) $r$ that contains some of the symbols, $0, 1, 2,m,e,n$. Any such  $r$ of length $t$ represents a certain set of length $t$ bitstrings. (This will be fully explained and motivated in later Sections). As a sneak preview, the number of length 10 bitstrings represented by $r=(2,m,e,m,1,n,e,e,1,n)$ is 84.  We say that $r$ is $\varphi$-{\it infeasible} with respect to a  10-variate Boolean function $\varphi$ if no bitstring in $r$ satisfies $\varphi$. Otherwise $r$ is called $\varphi$-{\it feasible}. If {\it all} bitstrings in $r$ satisfy  a Boolean formula $\psi$ then we say that $r$ {\it fulfills} $\psi$. 

{\bf 2.2} The input for the $men$-algorithm is any Boolean function $\varphi:\{0,1\}^t\to \{0,1\}$ given in CNF format $C_1\wedge C_2\wedge\ldots\wedge C_s$. The output of the men-algorithm is the {\it model set} $Mod(\varphi)$, i.e. the set of bitstrings ${\bf x}$ with $\varphi({\bf x})=1$. Here $Mod(\varphi)$ comes  as a disjoint union of 012men-rows. If there is no ambiguity we may simply speak of rows instead of 012men-rows.
The basic supporting data-structure is a Last-In-First-Out (LIFO)  stack,
 filled with changing 012men-rows. (It is well known that LIFO amounts to DFS=Depth-First-Search of a tree, but the author prefers the LIFO point of view.)
 At the beginning the only 012men-row in the LIFO stack is $(2,2,...,2)$, thus the powerset $\{0,1\}^t$, see (1). Suppose that by induction we obtained a LIFO stack as shown in Figure 1a (so each $*$ is one of the symbols $0,1,2,m,e,n$).

\includegraphics[scale=1.0]{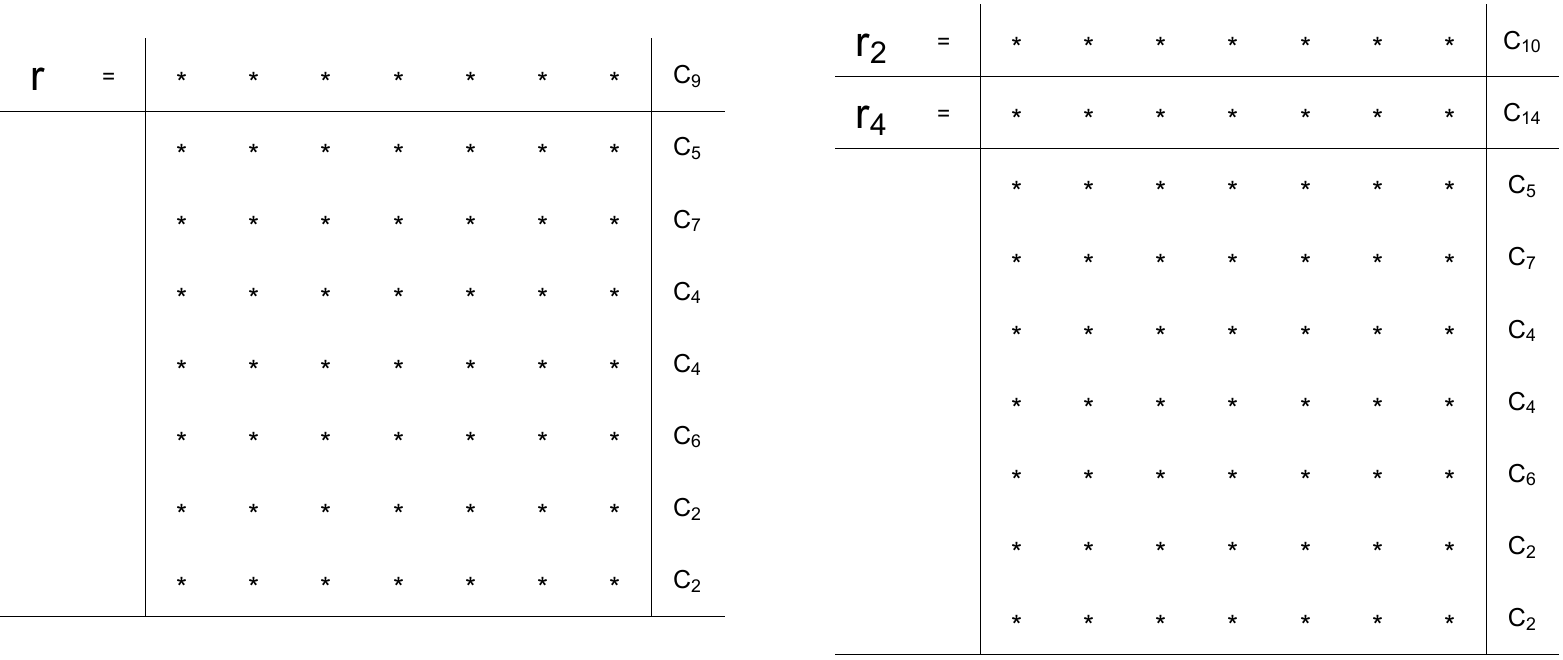}

{\sl Figure 1a: LIFO stack before imposing $C_9$ \hspace*{1.1cm} Figure 1b: LIFO stack after imposing $C_9$}

The top row $r$ fulfills $C_1\wedge ... \wedge C_8$, but not yet $C_9$, which hence is the {\it pending clause}. Similarly the other rows have pending clauses as indicated in the last column.
To {\it impose} $C_9$ upon $r$ means replacing $r$ by a few successor rows $r_i$, called the {\it sons} of $r$, whose union is disjoint and contains exactly those bitstrings in $r$ that satisfy $C_9$. This maneuver is the core novel ingredient of the $men$-algorithm (as opposed to  LIFO  or SAT-solvers which are present in every decent ALLSAT algorithm [TS]). Sections 3 to 6 deliver the details of how the sons $r_i$ get calculated. As shown in Section 5 the number of sons is bounded by the length of the imposed clause.

For now we illustrate the core maneuver with the Venn diagram in Figure 2. By assumption $r\subseteq Mod(C_1\wedge ... \wedge C_8)$ but
$r\not\subseteq Mod(C_9)$. The part $r\setminus Mod(C_9)$ 'melts away' and the remainder of $r$ gets decomposed into
 four {\it candidate sons} $r_1$ to $r_4$. Having discarded the $\varphi$-infeasible row $r_3$ (more details in 2.2.1) we turn to $r_1,\ r_2,\ r_4$. They all fullfil $C_9$ by consruction. Say $r_2$ does not fulfil $C_{10}$. Then its pending clause is $C_{10}$. Say $r_4$ happens to fulfill $C_{10}$ to $C_{13}$ but not $C_{14}$. Then its pending clause is $C_{14}$. Say $r_1$ happens to fulfill $C_1$ up to $C_s$. Then $r_1$ is {\it final} in the sense that $r_1\subseteq Mod(\varphi)$. One then removes $r_1$ from the LIFO stack and outputs (or stores) it as part of the required compressed delivery of $Mod(\varphi)$. The rows $r_2,\ r_4$ are the {\it sons} of $r$ and take its place (in any order)  on top of the LIFO stack, see Figure 1b. This finishes the imposition of $C_9$ upon $r$.

\includegraphics[scale=1.0]{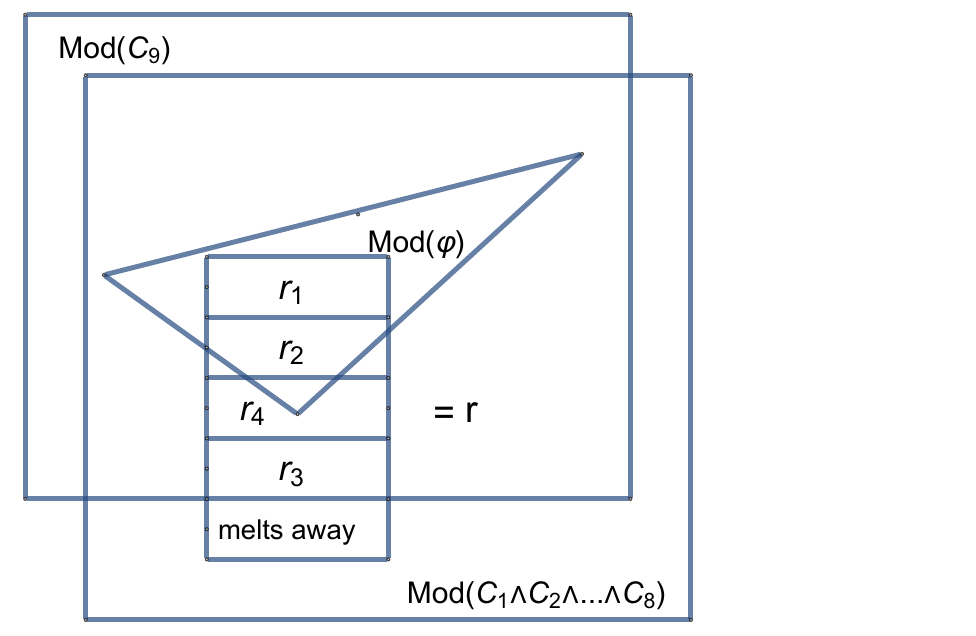}

{\sl Figure 2: Visualization of the core maneuver}

 {\bf 2.2.1.} That $r_3$ is $\varphi$-infeasible can be detected  
as follows. Translate $r_3$ into a Boolean CNF $\sigma$. (As a sneak preview, if $r_3=(e,0,e,1,e)$, then $\sigma\ =\ (x_1\vee x_3\vee x_5)\wedge \ol{x_2}\wedge x_4$.) Evidently $r_3$ is
$\varphi$-infeasible, if and only if $\varphi\wedge\sigma$ is insatisfiable. This can be determined with any off-the-shelf
 SAT-solver.
In contrast, determining the pending clause of a  row works  fast because for {\it any} 012men-row $r$ and  {\it any} given clause $C$ it is 
straightforward (Section 7) to check whether or not $r'$ fulfills $C$. 

{\bf 2.3} By induction at all stages the union $U$ of all final rows and of all rows in the LIFO stack is  disjoint and contains $Mod(\varphi)$. Whenever the pending clause of any top row $r$ gets imposed on $r$, a nonempty part of $r$ melts away, and so the new set $U$ strictly shrinks. Hence the procedure ends in finite time. Specifically, once the LIFO stack becomes empty, the set $U$ equals the disjoint union of all final rows, which in turn equals  $Mod(\varphi)$. See Section 6 for carrying out all of this with a concrete Boolean function $\varphi$.

\section{The Flag of Bosnia and its higher level variants}

{\bf 3.1} The real Flag of Bosnia\footnote{Strictly speaking Bosnia should be Bosnia-Herzegowina, but this long name gets too clumsy.  Other national flags, such as the Flag of Papua (used in previous publications), have  similar patterns but miss out on relevant details.} (FoB) features a white main diagonal, the lower triangle is blue, and the upper triangle yellow. Using $0,1,2$  as colors the two kinds of FoBes we care about are rendered in Figure 3 and 4. Here {\it Type} 1 and  {\it Type} 0 refers to the color of  the diagonal. 


\includegraphics[scale=0.58]{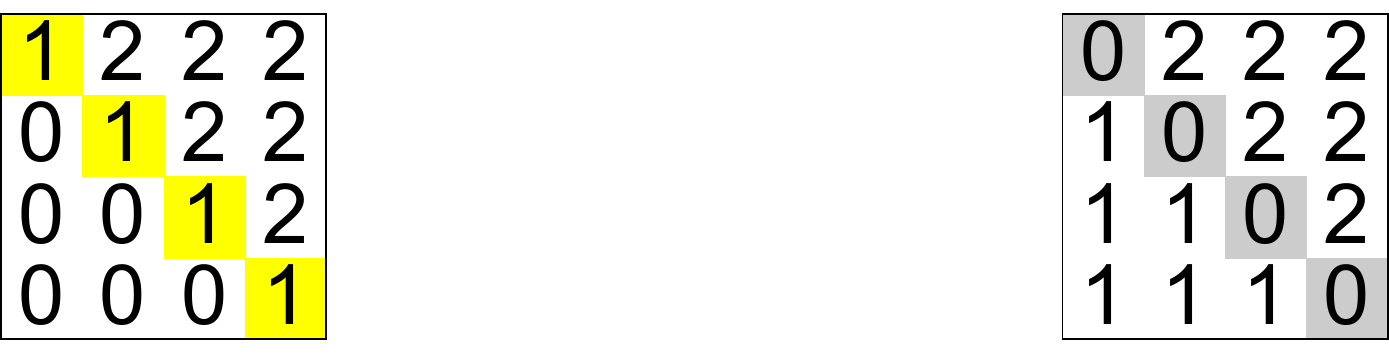}

{\sl Figure 3: FoB of Type 1 \hspace*{1.7cm} Figure 4: FoB of Type 0.}

The FoB of Type 1 visualizes in obvious ways the righthand side of the well-known tautology

(2)\qquad $(x_1\vee x_2\vee x_3\vee x_4)\ \leftrightarrow\  x_1\,\vee\,(\ol{x_1}\wedge x_2)\,\vee\,
(\ol{x_1}\wedge\ol{x_2}\wedge x_3)\,\vee\,(\ol{x_1}\wedge\ol{x_2}\wedge\ol{x_3}\wedge x_4)$

The dimension $4\times 4$ generalizes to any $k\times k$, but only $k\ge 2$ will be relevant. It is essential that the four clauses on the right in $(2)$ are mutually disjoint, i.e. their conjunction is insatisfiable.  Equation (2) (for any $k\ge 2$) is the key for many methods that {\it orthogonalize} an arbitrary DNF into an  {\it exclusive sums of products (ESOP)}; see [B,p.327]. It will be essential for us as well, but we orthogonalize CNF's, not DNF's. What is more, our use of wildcards results into 'fancy kinds of ESOPs', i.e. disjoint unions of 012men-rows.

As in previous publications we prefer to write 2 for the common don't-care symbol $\ast$. Thus  the $012$-{\it row} $(2,0,1,2,1)$ by definition is the set of bitstrings 

$\{({\bf 0},0,1,{\bf 0},1),({\bf 0},0,1,{\bf 1},1),({\bf 1},0,1,{\bf 0 },1),({\bf 1},0,1,{\bf 1},1)\}$.

Thus in view of (2) the  model set of $x_1\vee x_2\vee x_3\vee x_4$ is the disjoint union of the four $012$-rows constituting the FoB in Figure 3. This matches the row-wise cardinality count: $8+4+2+1=2^4-1$. 
Dually the FoB of Type 0    in Figure 4 visualizes the tautology

(3)\qquad $(\ol{x_1}\vee \ol{x_2}\vee \ol{x_3}\vee \ol{x_4})\ \leftrightarrow\ 
 \ol{x_1}\vee(x_1\wedge \ol{x_2})\vee(x_1\wedge x_2\wedge \ol{x_3})\vee
(x_1\wedge x_2\wedge x_3\wedge \ol{x_4})$

{\bf 3.2} More original than writing 2 instead of $\ast$, is it to dismiss the whole FoB in Figure 3 and replace it by the single wildcard $(e,e,e,e)$ which by {\it definition}\footnote{Surprisingly, this idea seems to be new. Information to the contrary is welcome. The definition generalizes to tuplets $(e,e...,e)$ of length $t\ge 2$. For simplicity we sometimes strip $(e,e,...,e)$ to $ee...e$. Observe that a single $e$ (which we forbid) would amount to $1$.} is the set of all length 4 bitstrings ${\bf x}=(x_1,x_2,x_3,x_4)$ with 'at least one 1'. In other words, only (0,0,0,0) is forbidden. Thus e.g $(1,e,0,e)$ is the set of bitstrings $\{(1,{\bf 1},0,{\bf 0}),\ (1,{\bf 0},0,{\bf 1}),\ (1,{\bf 1},0,{\bf 1})\}$. If several $e$-wildcards occur, they need to be distinguished by subscripts. For instance the $012e$-{\it row} $r_1$ in Figure 5a represents the model set of the CNF

(4)\qquad $(x_1\vee x_2\vee x_3\vee x_4)\wedge(x_5\vee x_6\vee x_7\vee x_8)$.

The $e$ symbols need {\it not be contiguous}. 
 But for better visualization our examples tend to clump $e$-symbols with the same subscript. 
Not all symbols $0,1,2,e$ need to occur in a $012e$-row. In other words, $012$-rows are special cases of $012e$-rows.

{\bf 3.3} The fewest number of disjoint $012$-rows required to represent the single row $r_1$ seems to be a hefty sixteen. These 012-rows are obtained by 'multiplying out' two FoBes of Type 1. Thus the $e$-wildcard boosts compression. But can the $e$-formalism handle overlapping clauses? It is here where the FoBes dismissed in 3.2 get vindicated, but they need to reinvent themselves as 'Meta-FoBes'. To fix ideas, let 
${\cal F}:=Mod(C_1\wedge C_2\wedge C_3)\subseteq Mod(C_1\wedge C_2)=r_1$, where

(5)\qquad $C_1\wedge C_2\wedge C_3:=(x_1\vee x_2\vee x_3\vee x_4)\wedge(x_5\vee x_6\vee x_7\vee x_8)\wedge(x_3\vee x_4\vee x_5\vee x_6)$.

We claim that ${\cal F}$ is the disjoint union of the two $012e$-rows  $r_2$ and $r_3$ in Figure 5a, and shall refer to the framed part as a {\it Meta-FoB} (of dimensions $2\times 2$). Specifically, the bitstrings $(x_3,x_4,x_5,x_6)$ satisfying the overlapping clause
$x_3\vee x_4\vee x_5\vee x_6$ are collected in $(e,e,e,e)$ and come in two sorts. The ones with $x_3=1$ or $x_4=1$ are collected in $(e,e,2,2)$, and the other ones are in $(0,0,e,e)$. These two quadruplets constitute, up to some adjustments, the two rows of our Meta-FoB. 

The first adjustment is that the right half of $(e,e,2,2)$ gets erased by the left part of the old constraint $(e_2,e_2,e_2,e_2)$ in $r_1$. The further adjustments do not concern the shape of the Meta-FoB per se, but rather are {\it repercussions} caused by the Meta-FoB outside of it. Namely, $(e_1,e_1,e_1,e_1)$ in $r_1$ splits into $(2,2,e,e)$ (left half of $r_2$) and $(e_1,e_1,0,0)$ (left part of $r_3$). It should be clear why $(e_2,e_2,e_2,e_2)$ in $r_1$ transforms differently: It stays the same in $r_2$ (as noticed already), and it becomes $(e,e,2,2)$ in $r_3$.
Because of its diagonal entries (shaded) our Meta-FoB  is\footnote{Generally the lengths of the diagonal entries $e...e$ match the cardinalities of the traces of the overlapping clause. For instance, imposing $x_4\vee x_5$ instead of $x_3\vee x_4\vee x_5\vee x_6$ triggers the Meta-FoB of Type $(e,1)$ in Figure 5b. We keep the terminology Type $(e,1)$ despite the fact that all diagonal entries are $1$. Confusion with FoBes of Type 1 (Figure 3)is unlikely.} a Meta-FoB of {\it Type $(e,1)$}. Why defining $r_2$ and $r_3$ in such complicated ways? Isn't $r_2\uplus r_3$ just the same as $r_1\setminus (e_1,e_1,0,0,0,0,e_2,e_2)$? Yes it is (and it matches 180+36=225-9), but the $men$-algorithm only digests set systems rendered as disjoint set-unions, it cannot handle set-differences.

\includegraphics[scale=0.46]{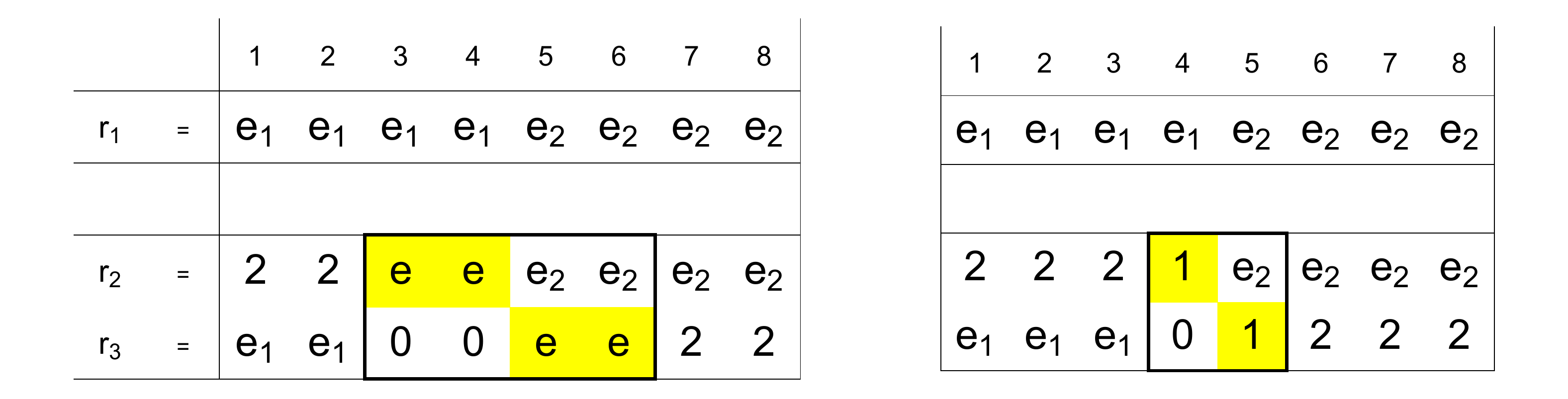}

{\sl \hspace*{0.5cm} Figure 5a:  Meta-FoB of Type $(e,1)$ \hspace*{2cm} Fig. 5b:  Small Meta-FoB of Type $(e,1)$ }

{\bf 3.4} In  dual fashion we define a second wildcard $(n,n,...,n)$ as the set of all length $t$ bitstrings  that have 'at least one 0' (where $t$ is the number of $n$'s).
We define $012n$-{\it rows} dually to 012e-rows. Mutatis mutandis the same arguments as above show that by using a dual Meta-FoB of Type $(n,0)$ one can impose $(n,n,...,n)$ upon disjoint constraints $(n_i,n_i,...,n_i)$. See Figure 6 which shows that the model set of

(5')\qquad $(\ol{x_1}\vee\ol{x_2} \vee \ol{x_3}\vee \ol{x_4})\wedge(\ol{x_5}\vee\ol{x_6} \vee \ol{x_7}\vee \ol{x_8})\wedge(\ol{x_3}\vee\ol{x_4} \vee \ol{x_5}\vee \ol{x_6})$

can be represented as disjoint union of the two $012n$-rows $r_2$ and $r_3$.

\includegraphics[scale=0.49]{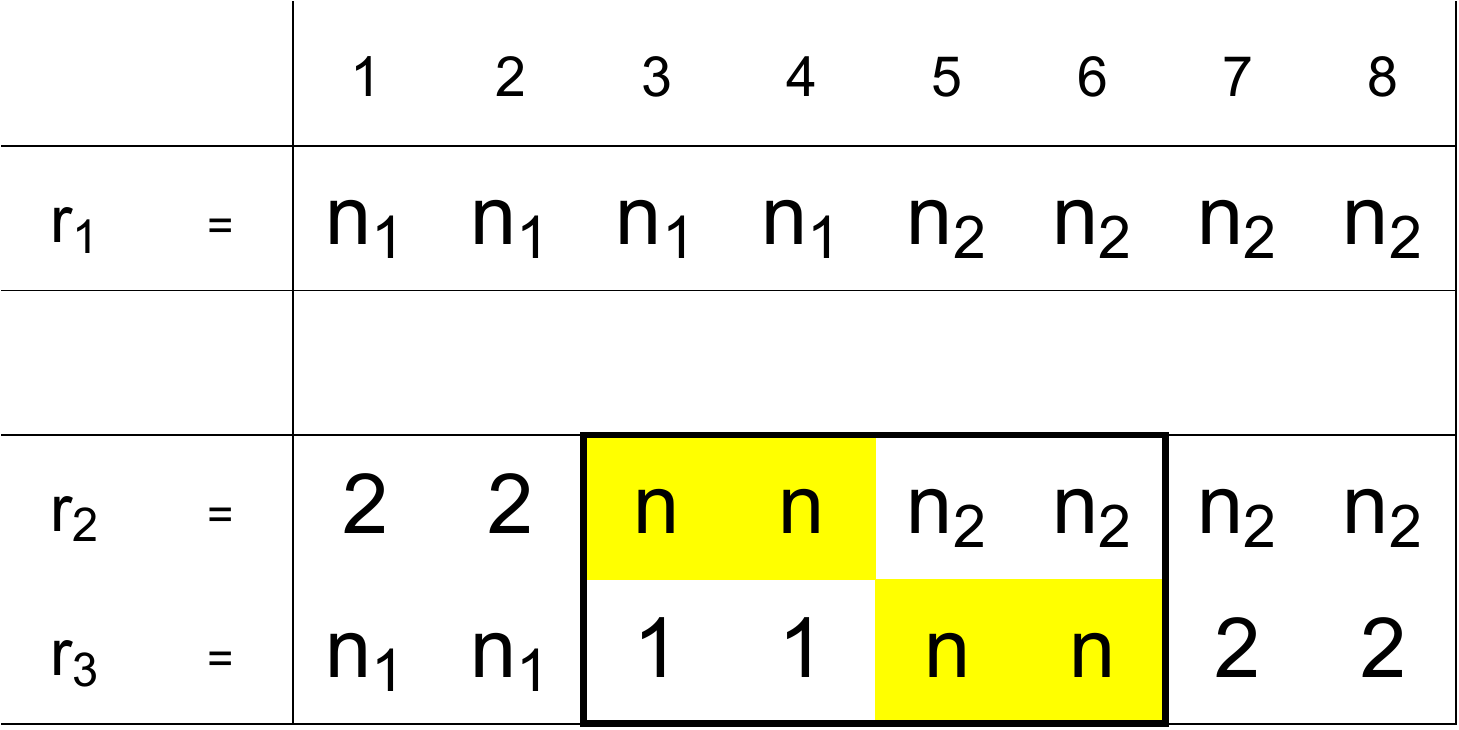}

{\sl Figure 6: Meta-FoB of Type $(n,0)$}

\section{Positive and negative clauses simultaneously}

New issues arise if $nnnn$ (or dually $eeee$) needs to be imposed on {\it distinct} types of wildcards, say $n_1n_1n_1n_1$ and $e_1e_1e_1e_1$ as occuring in row $r_1$ of Figure 7.
Specifically, let $n_1n_1n_1n_1$ model $C_1\,=\,\ol{x_1}\vee\ol{x_2}\vee\ol{x_3}\vee\ol{x_4}$, let $e_1e_1e_1e_1$ model $C_2=x_5\vee x_6\vee x_7\vee x_8$, and
$nnnn$ model the overlapping clause $C_3\,=\,\ol{x_3}\vee\ol{x_4}\vee\ol{x_5}\vee\ol{x_6}$.
 We need to sieve the model set ${\cal F}:=Mod(C_1\wedge C_2\wedge C_3)$ from  $r_1:=Mod(C_1\wedge C_2)$. Thus we need to represent $\{{\bf x}\in r_1:\ {\bf x}\ {\it satisfies}\ C_3\}$ in compact format. To do so write
 $r_1=r_2\uplus r_2'$, 
where 

$r_2:=\{{\bf x}\in r_1:\ {\bf x}\ {\it satisfies}\ \ol{x_3}\vee\ol{x_4}\}\ =\ \{{\bf x}\in r_1:\ x_3=0\,{\it or}\, x_4=0\}$, 

$r_2':=\{{\bf x}\in r_1:\ {\bf x}\ {\it violates}\ \ol{x_3}\vee \ol{x_4}\}\ =\ \{{\bf x}\in r_1:\ x_3=x_4=1\}$. 

It follows that $r_2\subseteq {\cal F}$ and in fact

(6)\qquad ${\cal F}\ =\ r_2\uplus \{{\bf x}\in r_2':\  {\bf x}\ {\it satisfies}\ C_3\}$

\includegraphics[scale=0.53]{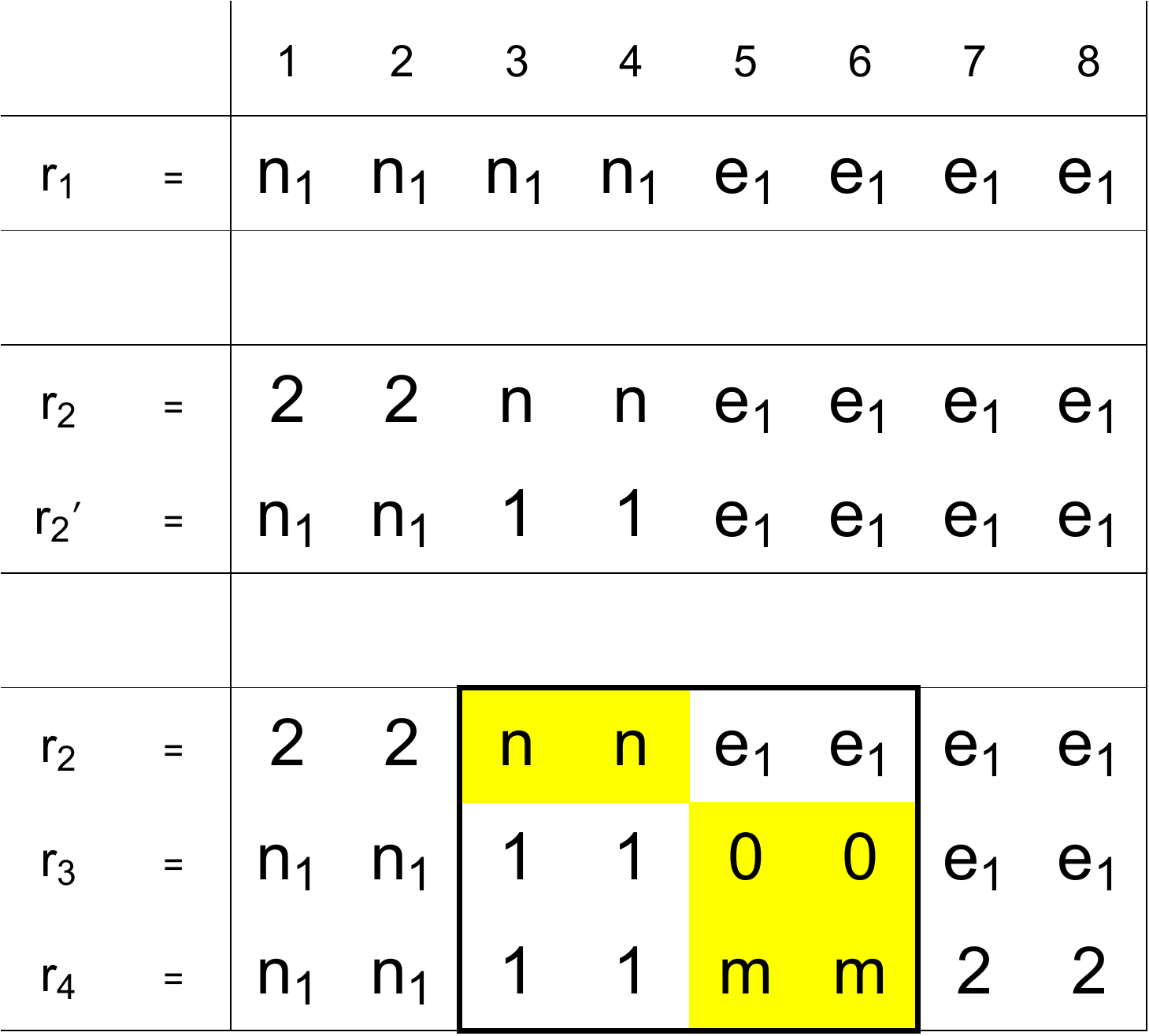}

{\sl Figure 7: Meta-FoB of Type $(n,m,0)$}

It is clear that $r_2$ and $r_2'$ can be rendered as in Figure 7.
For instance ${\bf x}=(0,0,1,1,1,1,1,1)$ is in $r_2'$ but does not satisfy $C_3$. How can one represent the rightmost set in (6) in a useful format?

To do so we define a third  wildcard 
$$(m,m,\ldots,m):=(e,e,\ldots,e)\cap(n,n,\ldots,n)$$ 
In other words, $(m,m,\ldots,m)$ is the set of all bitstrings with 'at least one  1 and at least one 0'.

A moment's thought  shows 
that the rightmost set in (6) is the disjoint union of $r_3$ and $r_4$ in Figure 7. The framed part in Figure 7 constitutes a  Meta-FoB
of {\it Type} $(n,m,0)$, i.e. all diagonal entries are $n$ or $m$ or $0$. 

 While $ee...e$ and $nn...n$ are duals of each other, $mm...m$ is selfdual. Hence in a dual way we can impose $ee...e$
 (matching $x_3\vee x_4\vee x_5\vee x_6$) upon $r_1$
by virtue of a Meta-FoB of {\it Type} $(e,m,1)$. This is carried out in Figure 8. 
We note in passing that the choice of letters $n$ and $m$ stems from 'nul' and 'mixed' 
respectively. The letter $e$ stems from 'eins' which is German for 'one'.

\includegraphics[scale=0.48]{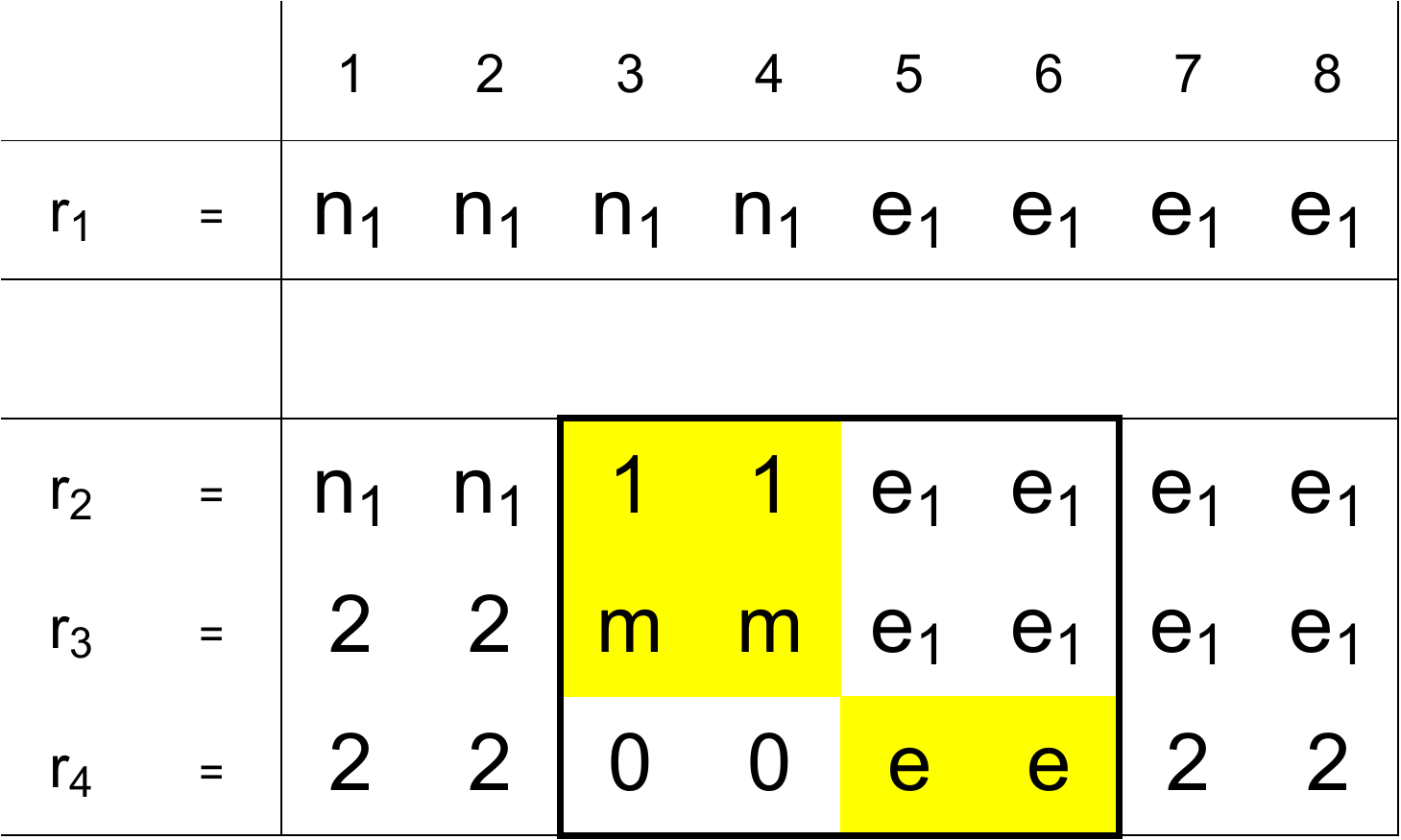}

{\sl Figure 8:  Meta-FoB of Type $(e,m,1)$}

\section{Imposing a positive or negative clause upon a 012men-row}

Having a third wildcard $mm...m$ proved to be useful in Section 4, but the prize is that we  need to cope with general $012men$-{\it rows} (defined in the obvious way) and
impose $nn...n$ or $ee...e$ (or even $mm...m$) upon them! Fortunately imposing $mm...m$ won't be necessary  and the imposition of $nn...n$ or $ee...e$ upon a 012men-row can be achieved using Meta-FoBes of Type $(n,m,0)$ and $(e,m,1)$ respectively. 

 In Figure 9 the imposition of 
$\ol{x_3}\vee\ol{x_4}\vee\ol{x_6}\vee\ol{x_7}\cdots\vee\ol{x_{14}} $  upon the $012men$-row $r_1$ is carried out (thus $nn...n$ has length 11 viewing that $\ol{x_5}$ is omitted). This boils down to the imposition of the shorter clause $\ol{x_6}\vee\ol{x_7}\cdots\vee\ol{x_{14}}$ since each ${\bf x}\in r_1$ has $x_3=x_4=1$. We omit the details of why the Meta-FoB of Type $(n,m,0)$, and its repercussions outside, look the way they look. For the most part this should be self-explanatory in view of our deliberations so far.

 \includegraphics[scale=0.77]{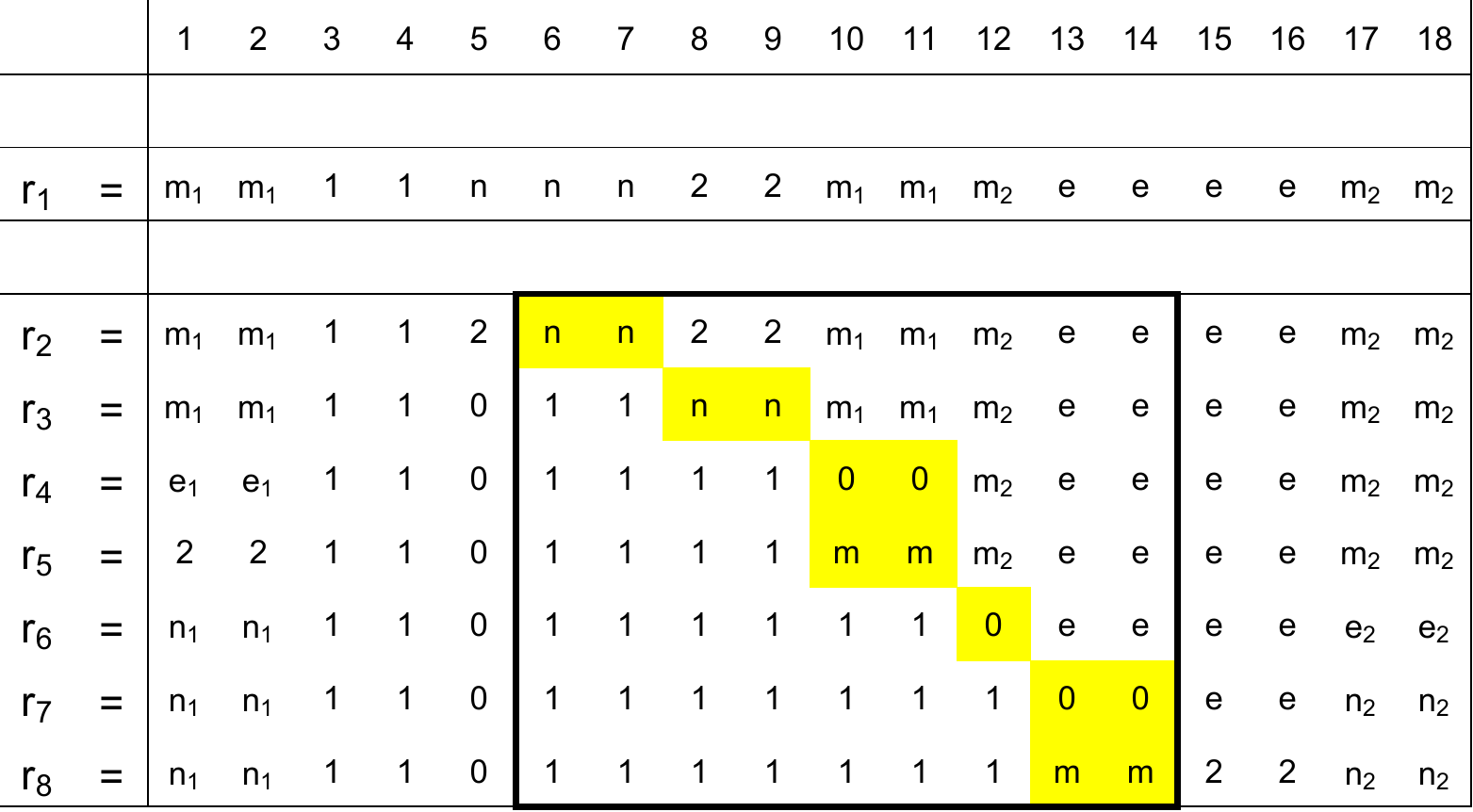}

{\sl Figure 9: Another Meta-FoB of Type $(n,m,0)$}

{\bf 5.1} But let us add a few comments in a different vein. Notice that $\ol{x_6}\vee\ol{x_7}\cdots\vee\ol{x_{14}}$ has 9 literals whereas the induced Meta-FoB has 7 rows. Generally speaking the shaded rectangles in any Meta-FoB arising from imposing a positive or negative clause upon $r$, are of dimensions $1\times t$ (any $t\ge 1$ can occur) and $2\times 2$. This implies that the number of rows in such a Meta-FoB (=number of sons of $r$) is at most the number of literals in that clause. Although imposing a {\it mixed} clause is more difficult (Section 6), it is easy to see that the number of literals remains an upper bound to the number of sons.

 {\bf 5.2} Imposing the corresponding {\it positive} clause  $x_3\vee x_4\vee\cdots\vee x_{14}$ upon $r_1$ would be trivial since each bitstring ${\bf x}$ in $r_1$ satisfies this clause in view of $x_3=x_4=1$. Energetic readers may enjoy imposing the shorter clause $ x_6\vee\cdots\vee x_{14}$ (thus $ee...e$) upon $r_1$ by virtue of a Meta-FoB of Type $(e,m,1)$.

In contrast, we dissuade imposing $mm...m$ upon $r_1$ by virtue of some novel Meta-FoB because for the time being\footnote{This concerns our present focus on arbitrary CNFs. For special types of CNFs, e.g. such that the presence of  $x_i\vee x_j\vee\cdots\vee x_k$ implies the presence of  $\ol{x_i}\vee \ol{x_j}\vee\cdots\vee\ol{x_k}$, imposing $mm...m$ may well be beneficial.}
only our capability (to be honed in Section 6) to
 either impose $ee...e$ or $nn...n$ matters. This skill, as well as a clever ad hoc maneuver, will suffice to impose any {\it mixed} clause upon any  012men-row.

\section{Handling general (=mixed) clauses}

 Let us  embark on the compression of the model set of the CNF with clauses

\begin{description}
\item{(7) }$\ C_1=\ol{x_1}\vee\ol{x_2}\vee\ol{x_3},\hspace{0.6cm} C_2=x_4\vee x_5\vee x_6\vee x_7,\hspace{0.6cm}  C_3=\ol{x_8}\vee\ol{x_9}\vee\ol{x_{10}}$,    
\item{} \hspace{0.6cm} $C_4=\ol{x_2}\vee\ol{x_3}\vee\ol{x_4}\vee\ol{x_5}\vee x_6\vee x_7\vee x_8\vee x_9,\hspace{0.6cm}  
          C_5=x_1\vee\ol{x_3}\vee\ol{x_4}\vee \ol{x_6}\vee\ol{x_7}$
\end{description}

It is clear that $r_1$ in Figure 10 compresses the model set of $C_1\wedge C_2\wedge C_3$. Hence the pending clause of $r_1$ is $C_4$. In order to sieve
 ${\cal F}:=Mod(C_1\wedge C_2\wedge C_3\wedge C_4)$ from $r_1=Mod(C_1\wedge C_2\wedge C_3)$ we first split $r_1$ as $r_1=r_2\uplus r_2'$ where

\begin{description}
\item{(8) } $r_2:=\,\{{\bf x}\in r_1:\ {\bf x}\ {\it satisfies}\ \ol{x_2}\vee\ol{x_3}\vee\ol{x_4}\vee\ol{x_5}\ \}\ $ {\it and}  
\item{} \hspace{0.7cm} $r_2':=\,\{{\bf x}\in r_1:\ {\bf x}\ {\it violates}\ \ol{x_2}\vee\ol{x_3}\vee\ol{x_4}\vee\ol{x_5}\ \}$.
\end{description}

Then we have $r_2\subseteq {\cal F}$, and  (akin to (6)) in fact

$(9)\quad {\cal F}=r_2\uplus \{{\bf x}\in r_2':\ {\bf x}\ {\it satisfies}\ x_6\vee x_7\vee x_8\vee x_9\ \}$.  

Similar to (6), but more demanding, both parts on the right in (9) must now be rewritten
 as disjoint union of $012men$-rows.

{\bf 6.1} Enter the 'ad hoc maneuver' mentioned above: Roughly speaking both bitstring systems $r_2$ and $r_2'$ temporarily morph into 'overloaded'  012men-rows. The latter will morph back,
  one after the other in 6.1.2 and 6.1.3,  in disjoint collections of (ordinary) 012men-rows.

Two definitions are in order. If in a 012men-row $r$ we bar any symbols, then the obtained {\it overloaded Type A row} by definition consists of the bitstrings in $r$ that feature at least one $0$ on a barred location. It follows that $r_2$ equals the overloaded Type A row with the same name in Figure 10. Similarly, if in a row $r$ we encircle, respectively decorate with stars, nonempty disjoint sets of  symbols, then the obtained {\it overloaded Type B row}  by definition consists of the bitstrings in $r$ that feature 1's at all encircled locations, and feature at least one 1 on the starred locations. It  follows that the rightmost set in (9) equals the overloaded Type B row $r_3$ in Figure 10.

We shall see that merely starring symbols (omitting encircling) also comes up. The definition of such an {\it overloaded Type C row} is as expected.

{\bf 6.1.2} As to turning $r_2$ and $r_3$ into ordinary $012men$-rows, we first look at $r_2$, while carrying along the overloaded row $r_3$. Transforming $r_2$ simply amounts to impose the negative part  $\ol{x_2}\vee\ol{x_3}\vee\ol{x_4}\vee\ol{x_5}$ of clause $C_4$ upon $r_1$, and hence works with the Meta-FoB of Type $(n,m,0)$ that stretches over $r_4$ to $r_6$. As to $r_5$, it fulfills $C_5$ (since each ${\bf x}\in C_5$ has $x_4=0$), and so is final and leaves the LIFO stack (Section 2).

{\bf 6.1.3} As to transforming $r_3$, the first step is to replace the encircled symbols by 1's and to record the ensuing repercussions. Some starred symbols may change in the process but they must {\it keep their star}. The resulting overloaded Type C row still represents the {\it same} set of bitstrings $r_3$. The second step is to 
impose the positive part $x_6\vee x_7\vee x_8\vee x_9$ of $C_4$ by virtue of a Meta-FoB, see $r_7$ to $r_9$ in Figure 10.

\includegraphics[scale=0.77]{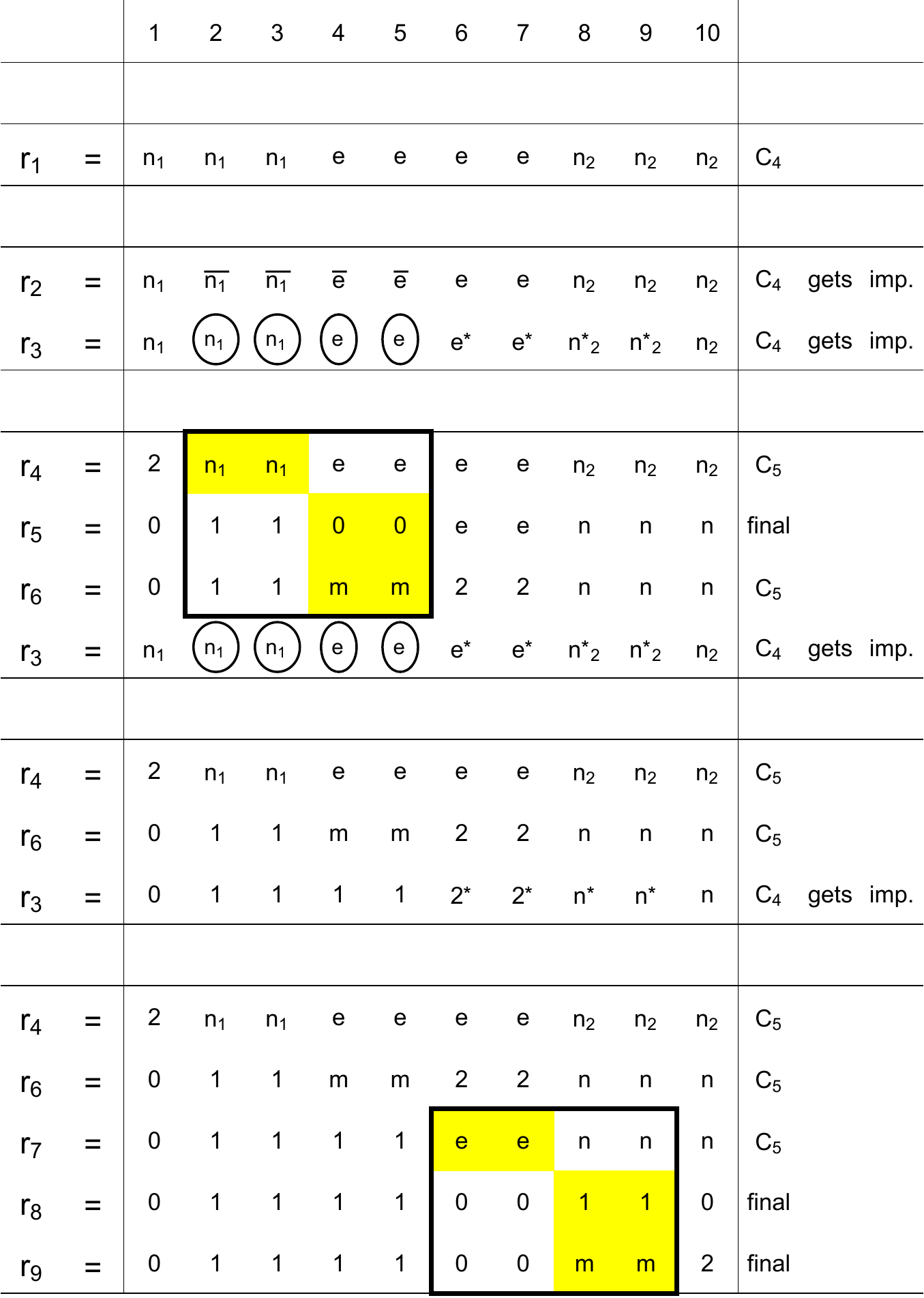} 

{\sl Figure 10: The $men$-algorithm in action. Snapshots of the LIFO stack.}

 {\bf 6.1.4} In likewise fashion (details left to the reader) the algorithm proceeds in Figure 11. Observe that in Figure 11 we {\it permuted} the columns in order to better visualize the imposition of clause $C_5$. Note that $r_{10},\ r_{11}$ are overloaded rows of Type A and B. The $men$-algorithm ends after the last row in the LIFO stack gets removed.
	
		\includegraphics[scale=0.77]{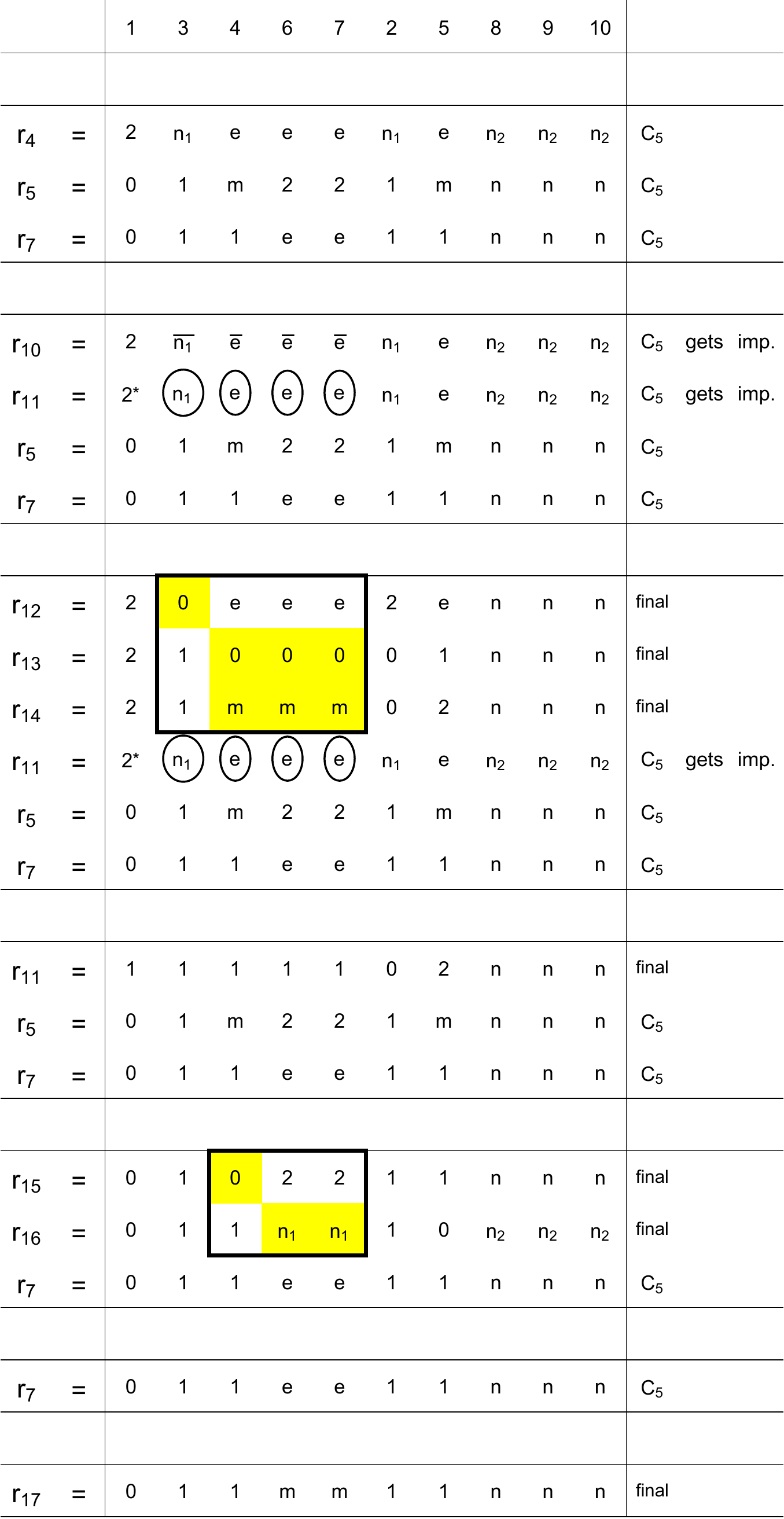} 
		
		{\sl Figure 11: Further snapshots of the LIFO stack.}
	
	Altogether there are ten (disjoint) final rows 
	$r_5,\ r_8,\ r_9,\ r_{12},\ r_{13},\ r_{14},\ r_{11},\ r_{15},\ r_{16},\ r_{17}$. Their union is  $Mod(\varphi)$, which hence is of cardinality
	
	$$|Mod(\varphi)|=21+1+4+420+14+168+14+28+21+14=695$$

\section{Testing whether a 012men-row fulfills a clause}

Here we verify the claim made in 2.2 that checking whether a 012men-row $r$ fulfills a clause $C$ is straightforward. Indeed, focusing on the most elaborate case of a {\it mixed} clause $C$ the following holds.

\begin{description}
\item[(10)] If $C=x_1\vee\cdots\vee x_s\vee \ol{x_{s+1}}\vee\cdots\vee \ol{x_t}$ and $r=(a_1,..,a_s,a_{s+1},..,a_t,\ldots)$ then $r$ fulfills $C$ iff one of these cases occurs:
\begin{description}
\item[(i)] For some $1\le j\le s$ one has $a_j=1$;
\item[(ii)] $\{1,\ldots,s\}$ contains the position-set of a full $e$-wildcard or full $m$-wildcard;
\item[(iii)] For some $s+1\le j\le t$ one has $a_j=0$;
\item[(iv)] $\{s+1,\ldots,t\}$ contains the position-set of a full $n$-wildcard or full $m$-wildcard;
\end{description}
\end{description}

{\it Proof of} $(10)$. It is evident that each of $(i)$ to $(iv)$ individually implies that all bitstrings ${\bf x}\in r$ satisfy $C$. Conversely suppose that $(i)$ to $(iv)$
are false. We must pinpoint a bitstring in $r$ that violates $C$. To fix ideas, consider  $r$ of length 18 and the clause $C=x_1\vee\cdots\vee x_6\vee \ol{x_7}\vee\cdots\vee\ol{x_{13}}$. (For readibility the disjunctions $\vee$ are omitted in Figure 12.)
Properties $(i)$ to $(iv)$ are false for $C$. For instance the position-set $\{6,7,8\}$ of $m_1m_1m_1$ is neither contained in $\{1,\ldots,s\}$ nor in $\{s+1,\ldots,t\}$. That it is contained in their union  is irrelevant.
One checks that $r_{\rm vio}\subseteq r$ and that each bitstring $x\in r_{\rm vio}$ violates $C$.

\includegraphics[scale=0.73]{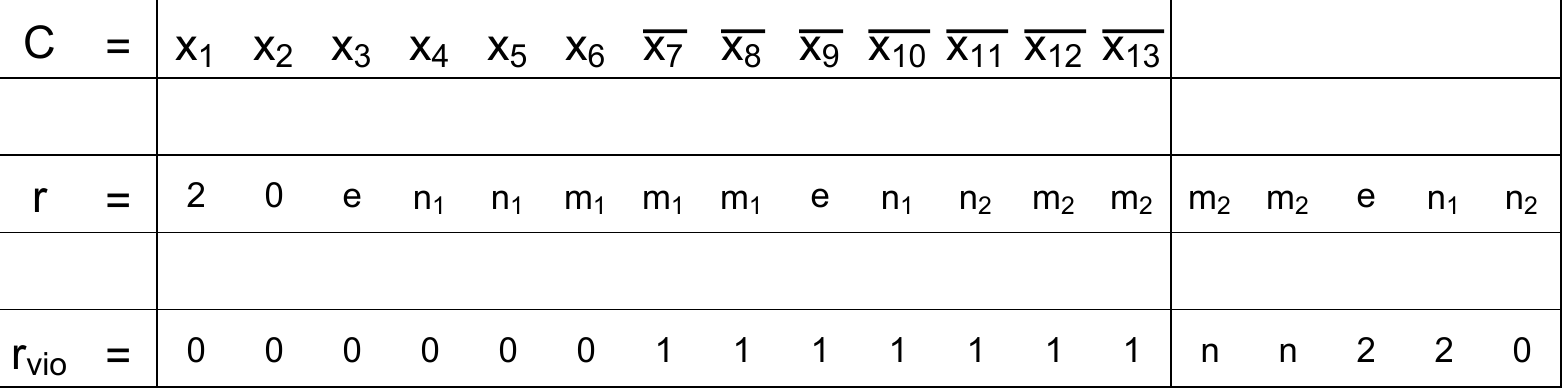} 

{\sl Figure 12: The $012men$-row $r$ does not fulfill clause $C$.}

\section{ Comparison with BDD's and ESOP's}

We reiterate from Section 1 that the $men$-algorithm has not yet been implemented. Therefore we content ourselves to take two medium-size random CNFs and hand-calculate what the $men$-algorithm does with them. We compare the outcome with two competing paradigms; ESOP's in 8.2, and BDD's in 8.3. But first we warm up in 8.1 by looking how ESOP and BDD handle $Mod(\mu_t)$ for $\mu_t=(x_1\vee\cdots\vee x_t)\wedge(\ol{x_1}\vee\cdots\vee\ol{x_t})$.
Recall that the $men$-algorithm achieves optimal compression here: $Mod(\mu_t)=(m,m,\ldots,m)$.

{\bf 8.1} One checks that the 012-rows of the Table on the right of Figure 13 constitute an ESOP of $\mu_5$. Let us verify that the BDD on the left in Figure 13 also yields $\mu_5$. As for any BDD, each nonleaf node A yields 'its own' Boolean function (on a subset of the variables).  For instance, there are two nodes labelled with $x_2$. The left, call it A, yields a Boolean function $\alpha(x_2,x_3,x_4,x_5)$ whose model set is the disjoint union of the four 012-rows in the top square in the Table on the right. For instance, the bitstring $(0,0,1,0)$ belongs to $(0,0,1,2)$,
and indeed it triggers (in the usual way, [K]) a path that leads from A to $\top$. Similarly the right node labelled $x_2$, call it B, yields some Boolean function $\beta(x_2,x_3,x_4,x_5)$
 whose model set is the disjoint union of the four 012-rows in the bottom square in the Table on the right. It is now evident that whole Table represents the model set of the whole BDD, thus $Mod(\mu_5)$.

Conversely, as is well known [B,p.327], each BDD gives rise\footnote{Unfortunately Mathematica does not openly support BDDs, and so the author had to turn to Python for that purpose. In general the Python BDD's yielded quite different ESOPs than Mathematica's ESOP-command. In the unlikely case that the latter is based on BDD's (I didn't manage to find out) this must be due to different variable orderings.} to an ESOP. It is easy to calculate its exclusive products, and even easier to to predict their number. See [W] for details.

\includegraphics[scale=1.1]{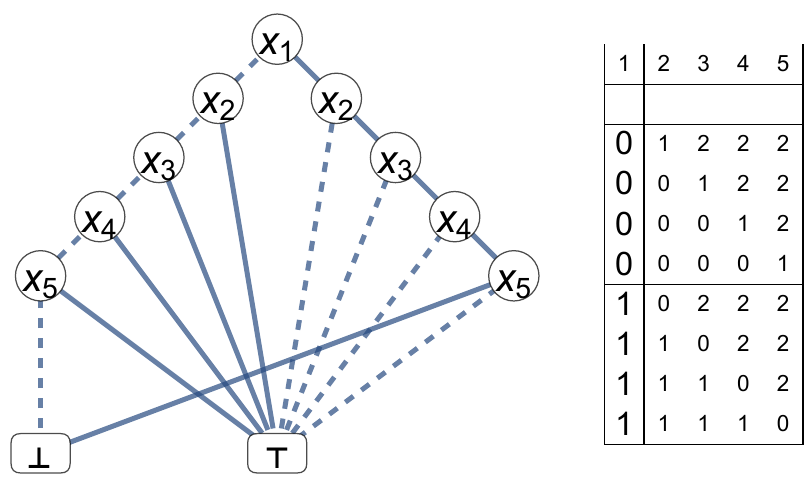}

 {\sl Figure 13: Each BDD readily yields an ESOP}

{\bf 8.2} Consider this CNF:

\begin{description}
\item{(11)} $\ \varphi_1=(x_5\vee x_7\vee x_{10}\vee\ol{x_2}\vee\ol{x_4})\wedge(x_1\vee x_2\vee x_9\vee\ol{x_7}\vee\ol{x_5})$
   
\item{} \hspace{1.6cm} $\wedge\ 
(x_2\vee x_3\vee x_7\vee\ol{x_4}\vee\ol{x_9})\wedge(x_8\vee x_9\vee x_{10}\vee\ol{x_4}\vee\ol{x_9})$
\end{description}

All clauses have 3 positive and 2 negative literals, which were randomly chosen (but avoiding $x_i\vee\ol{x_i}$) from a set of 20 literals. Table 14 shows the fourteen rows that the men-algorithm produces to compresses $Mod(\varphi_1)$. One reads off that $|Mod(\varphi_1)|=16+48+\cdots+18=898$.

\includegraphics[scale=0.83]{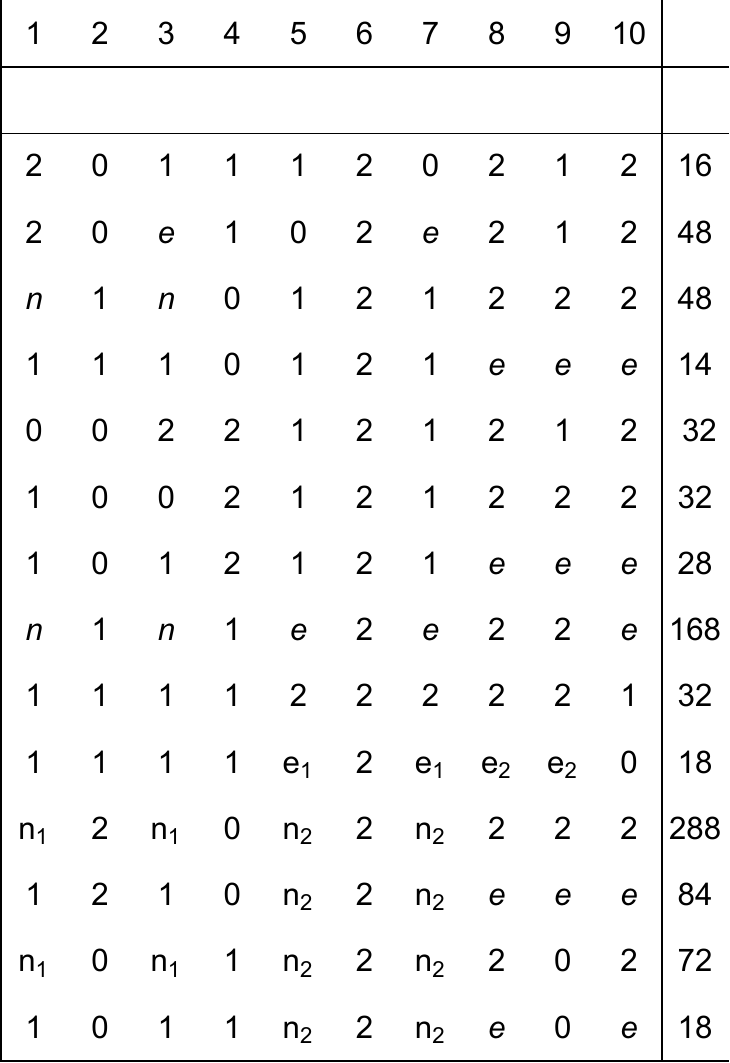} 

{\sl Table 14: Applying the men-algorithm to $\varphi_1$ in (11).}

Using the Mathematica-command {\tt BooleanConvert} (option "ESOP") transforms (11)  to an ESOP $(\ol{x_4}\wedge x_9)\vee(x_1\wedge\ol{x_4}\wedge x_8\wedge\ol{ x_9})\vee\cdots$, which amounts to a union $(2,2,2,0,2,2,2,2,1,2)\cup\hfill$ 
$(1,2,2,0,2,2,2,1,0,2)\cup\cdots$ of 23 disjoint 012-rows. We note that the ESOP algorithm is quite sensitive\footnote{And so is the men-algorithm. For both methods, no attempt to optimize clause order has been made.} to the order of clauses. Incidentally the 23 rows above stem from one of the optimal permutations of clauses; the worst would yield 36 rows. Adding the random clause $(x_5\vee x_6\vee x_8\vee\ol{x_3}\vee\ol{x_9})$ to $\varphi_1$ triggers twenty six 012men-rows, but between 27 and 56 many 012-rows
with the ESOP-algorithm.

The second example in (12) has longer clauses, all of them either positive or negative (for ease of hand-calculation). Long clauses make our wildcards more effective still.

\begin{description}
\item{(12)} $\varphi_2=(x_3\vee x_4\vee x_6\vee x_7\vee x_9\vee x_{14}\vee x_{15}\vee x_{16}\vee x_{17}\vee x_{18})$  
\item{} \hspace{1.4cm} $\wedge\ (\ol{x_3}\vee\ol{x_5}\vee\ol{x_8}\vee\ol{x_9}\vee\ol{x_{11}}\vee\ol{x_{12}}\vee\ol{x_{13}}\vee\ol{x_{14}}\vee\ol{x_{15}}\vee\ol{x_{17}})$
\item{} \hspace{1.4cm} $\wedge\ (x_1\vee x_4\vee x_5\vee x_6\vee x_9\vee x_{12}\vee x_{14}\vee x_{15}\vee x_{17}\vee x_{18}) $
\item{} \hspace{1.4cm} $\wedge\ (\ol{x_1}\vee\ol{x_2}\vee\ol{x_3}\vee\ol{x_8}\vee\ol{x_{11}}\vee\ol{x_{13}}\vee\ol{x_{14}}\vee\ol{x_{16}}\vee\ol{x_{17}}\vee\ol{x_{18}})$
\item{} \hspace{1.4cm} $\wedge\ (x_2\vee x_3\vee x_7\vee x_8\vee x_{11}\vee x_{13}\vee x_{14}\vee x_{16}\vee x_{17}\vee x_{18}) $
\end{description}

Table 15 shows the ten rows the men-algorithm uses to compress $Mod(\varphi_2)$. In contrast the ESOP-algorithm uses between 85 and 168 many 012-rows, depending on the order of the clauses.

\includegraphics[scale=1.04]{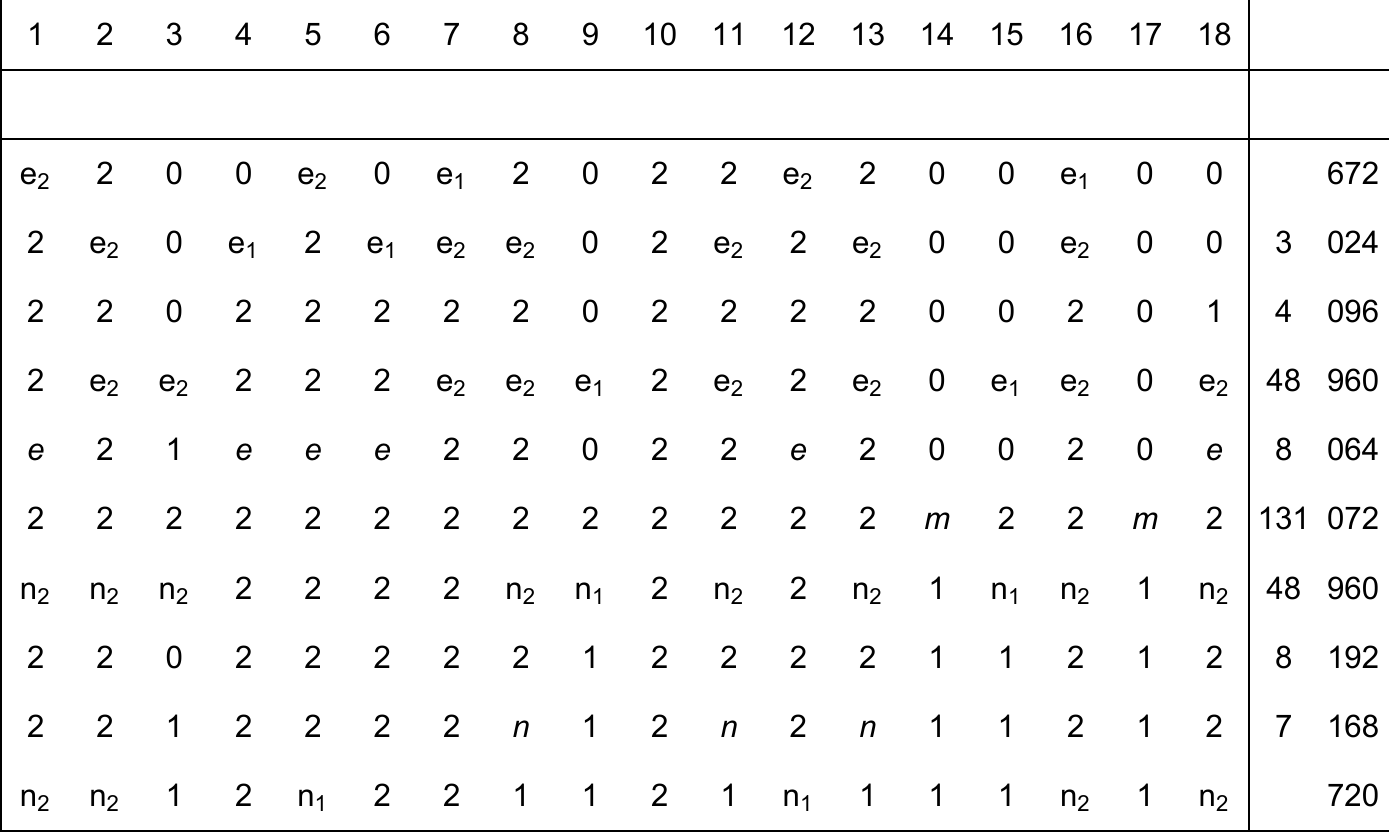} 

{\sl Table 15: Applying the men-algorithm to $\varphi_2$ in (12).}

{\bf 8.3} As to BDD's, one of many\footnote{Recall that the size of a BDD greatly depends on the chosen variable order. The variable order can be optimized in intelligent ways [K] but that costs time. The author does not know whether Python 3.5.2 embarks on such manoevers.} BDD's of $\varphi_2$ is rendered in Figure 16 below. It has 60 nodes and induces (in the way sketched in 8.1) an ESOP with 173 exclusive products.

\includegraphics[scale=0.99]{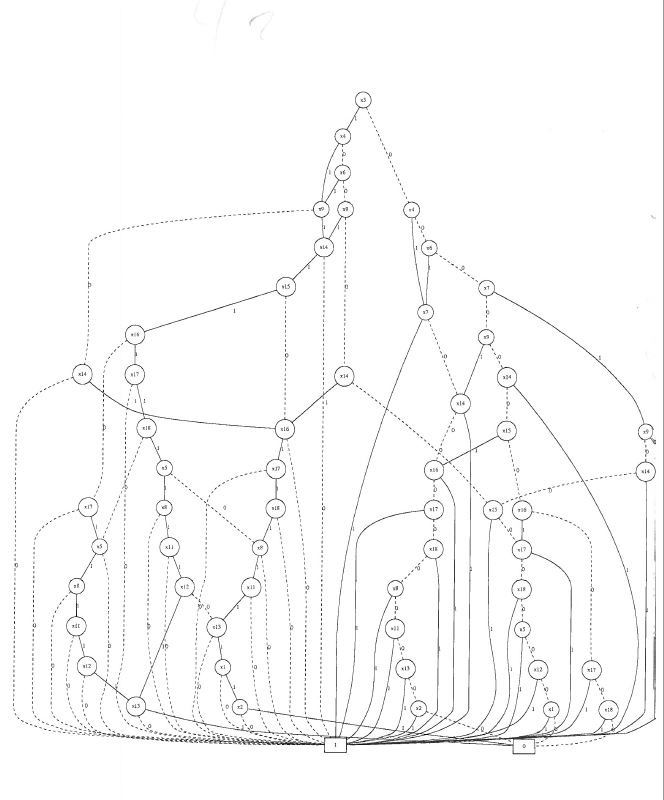}

{\sl Figure 16: Some BDD of $\varphi_2$.}

\section*{References}
\begin{enumerate}
\item[{[B]}] E. Boros, Orthogonal forms and shellability, Section 7 in: Boolean Fuctions (ed. Y. Crama, P.L. Hammer), Enc. Math. Appl. 142, Cambridge University Press 2011.
\item[{[K]}] D. Knuth, The art of computer programming, Volume 4 (Preprint), Section 7.14: Binary decision diagrams, Addison-Wesley 2008.
	\item [{[TS]}] Takahisa Toda and Takehide Soh. 2016. Implementing Efficient All Solutions SAT Solvers.
	J. Exp. Algorithmics 21, Article 1.12 (2016), 44 pages. DOI: https://doi.org/10.1145/2975585	
	\item[{[W]}] M. Wild, ALLSAT compressed with wildcards: Converting CNFs to orthogonal DNFs, ResearchGate. 
\end{enumerate}

\end{document}